\documentclass[prb,superscriptaddress,twocolumn]{revtex4-1}
\usepackage{amsmath,mathrsfs}
\usepackage{graphicx}
\usepackage{rotating}
\usepackage{amssymb}

\usepackage[usenames, 
            dvipsnames
            ]{color}

\usepackage[colorlinks=true,
            linkcolor=Blue,
            filecolor=magenta,
            citecolor=Blue,
            urlcolor=Blue,
            pdftitle={Dipolar Spins},
            pdfauthor={A. Keles},
            bookmarks=true
            ]{hyperref}

\usepackage[sort&compress]{natbib}

\usepackage{float}

\usepackage{flushend} 

\usepackage[numbered, 
            open
            ]{bookmark}

\usepackage[all]{hypcap}

\graphicspath{ {./figs/} }

\usepackage{listings}

\usepackage{tikz}

\tikzset{middlearrow/.style={
    decoration={markings,
      mark= at position 0.55 with {\arrow[scale=1,blue]{#1}} ,
    },
    postaction={decorate}
  }
}
\usetikzlibrary{decorations.pathreplacing,
  decorations.markings,decorations.pathmorphing,
    calc,
  shapes}

\usepackage{multirow}

\usepackage{pifont}

\newcommand\dgr{\ensuremath^\dagger}

\newcommand\x{\ensuremath\mathbf{x}}

\usepackage{bbold}

\newcommand\w{\omega}
\newcommand\al{\alpha}
\newcommand\pb{\mathbf{p}}

\newcommand\bdiamond{\tikz{\node[draw,scale=0.4,diamond,fill=black](){};}}

\newcommand{\oneloop}[1]
{
\begin{tikzpicture}[scale=.4,baseline=(current bounding box.center)]

    \def \x {0}
    \def \y {.1}
    \def \w {.3}
    \def \l {.5}  
    \def \dver {3}
    \draw[fill=gray] (\x,\y) rectangle (\x+\w,\y+\w);
    \draw[fill=gray] (\x+\w+\dver,\y) rectangle (\x+\dver+2*\w,\y+\w);
    \foreach \m/\n [count=\i] in {#1}
    {
      \ifthenelse{\equal{\i}{1}}
      {
        \ifx\m\n
          \draw[middlearrow={stealth}] 
        \else
          \draw[middlearrow={stealth reversed}] 
        \fi
        (\x-\l,\y+\w+\l) -- (\x,\y+\w);
        \node [left, black] at (\x-\l,\y+\w+\l) 
        {\tiny\ensuremath{\m}};
      }{}  
      \ifthenelse{\equal{\i}{2}}
      {
        \ifx\m\n
        \draw[middlearrow={stealth}] 
        \else
        \draw[middlearrow={stealth reversed}] 
        \fi
        (\x-\l,\y-\l) -- (\x,\y);
        \node [left, black] at (\x-\l,\y-\l) 
        {\tiny\ensuremath{\m}};
      }{} 
      \ifthenelse{\equal{\i}{3}}
      {
        \ifx\m\n
        \draw[middlearrow={stealth}] 
        \else
        \draw[middlearrow={stealth reversed}] 
        \fi
        (\x+2*\w+\dver,\y+\w) -- (\x+2*\w+\dver+\l,\y+\w+\l);
        \node [right, black] at (\x+2*\w+\dver+\l,\y+\w+\l) 
        {\tiny\ensuremath{\m}};
      }{} 
      \ifthenelse{\equal{\i}{4}}
      {
        \ifx\m\n
        \draw[middlearrow={stealth}] 
        \else
        \draw[middlearrow={stealth reversed}] 
        \fi
        (\x+2*\w+\dver,\y) -- (\x+2*\w+\dver+\l,\y-\l);
        \node [right, black] at (\x+2*\w+\dver+\l,\y-\l) 
        {\tiny\ensuremath{\m}};
      }{} 
      \ifthenelse{\equal{\i}{5}}
      {
        \ifx\m\n
        \draw[middlearrow={stealth reversed}] 
        \else
        \draw[middlearrow={stealth }] 
        \fi
        (\x+\w+\dver,\y+\w) 
        to [out=145,in=35]
        (\x+\w,\y+\w);
        \node [above, black] at (\x+\w+.5*\dver,\y+\w+.5)
        {\tiny\ensuremath{\m}};
      }{} 
      \ifthenelse{\equal{\i}{6}}
      {
        \ifx\m\n
        \draw[middlearrow={stealth reversed}] 
        \else
        \draw[middlearrow={stealth }] 
        \fi
        (\x+\w+\dver,\y) 
        to [out=-145,in=-35] 
        (\x+\w,\y);
        \node [below, black] at (\x+\w+.5*\dver,\y-0.5) 
        {\tiny\ensuremath{\m}};
      }{} 
    } 
  \end{tikzpicture}
}

\def\iline(#1,#2,#3,#4,#5,#6)
{
  \draw[
  thick,
  decorate,
  decoration={snake, amplitude=0.3mm, segment length=1.0mm}
  ]
  (#1,#2) -- (#3,#4); 
}
\def\fline(#1,#2,#3,#4,#5,#6)
{
  \draw[middlearrow={stealth}] (#1,#2) to (#3,#4); 
} 
\newcommand{\vertex}[1]
{
\begin{tikzpicture}[scale=.3,baseline=(current bounding box.center)]

    \def \x {0}
    \def \y {.1}
    \def \w {.5}
    \def \l {.6}  
    \def \dver {3}
    \draw[fill=gray] (\x,\y) rectangle (\x+\w,\y+\w);
    \foreach \m/\n [count=\i] in {#1}
    {
      \ifthenelse{\equal{\i}{1}}
      {
        \ifx\m\n
          \draw[middlearrow={stealth}] 
        \else
          \draw[middlearrow={stealth reversed}] 
        \fi
        (\x-\l,\y+\w+\l) -- (\x,\y+\w);
        \node [left, black] at (\x-\l,\y+\w+\l) 
        {\tiny\ensuremath{\m}};
      }{}  
      \ifthenelse{\equal{\i}{2}}
      {
        \ifx\m\n
        \draw[middlearrow={stealth}] 
        \else
        \draw[middlearrow={stealth reversed}] 
        \fi
        (\x-\l,\y-\l) -- (\x,\y);
        \node [left, black] at (\x-\l,\y-\l) 
        {\tiny\ensuremath{\m}};
      }{} 
      \ifthenelse{\equal{\i}{3}}
      {
        \ifx\m\n
        \draw[middlearrow={stealth}] 
        \else
        \draw[middlearrow={stealth reversed}] 
        \fi
        (\x+\w,\y+\w) -- 
        (\x+\w+\l,\y+\w+\l);
        \node [right, black] at (\x+\w+\l,\y+\w+\l)
        {\tiny\ensuremath{\m}};
      }{} 
      \ifthenelse{\equal{\i}{4}}
      {
        \ifx\m\n
        \draw[middlearrow={stealth }] 
        \else
        \draw[middlearrow={stealth reversed}] 
        \fi
        (\x+\w,\y) -- 
        (\x+\w+\l,\y-\l);
        \node [right, black] at (\x+\w+\l,\y-\l) 
        {\tiny\ensuremath{\m}};
      }{} 
    } 
  \end{tikzpicture}
}
\newcommand{\oneloopV}[1]
{
  \begin{tikzpicture}[scale=.8, baseline=(current bounding box.center)]
    \ifthenelse{\equal{1}{#1}}
    {
      \iline(1,0,1,1,d,a )  
      \iline(2,0,2,1,d,a )  

      \fline(0.5,0,1,0,d,a )  
      \fline(1,0,2,0,d,a )  
      \fline(2,0,2.5,0,d,a )  

      \fline(0.5,1,1,1,d,a )  
      \fline(1,1,2,1,d,a )  
      \fline(2,1,2.5,1,d,a )
    }{}  
    \ifthenelse{\equal{2}{#1}}
    {
      \iline(1.5, 0.0, 1.5, 0.3, d, a )  
      \iline(1.5, 0.7, 1.5, 1.0, d, a )  

      \fline(0.5, 0.0, 1.5, 0.0, d, a )  
      \fline(1.5, 0.0, 2.5, 0.0, d, a )  

      \fline(0.5, 1.0, 1.5, 1.0, d, a )  
      \fline(1.5, 1.0, 2.5, 1.0, d, a )  
      \draw[middlearrow={stealth}] (1.5, 0.7) to[out=0,in=0, looseness=1.5]  (1.5,0.3); 
      \draw[middlearrow={stealth}] (1.5, 0.3) to[out=180,in=180, looseness=1.5]  (1.5,0.7);
    }{}  
    \ifthenelse{\equal{3}{#1}}
    {
      \fline(1,0,1.5,.5,d,a )  
      \fline(1.5,.5,2,0,d,a )  
      \iline(1.5,.5,1.5,1,d,a )  

      \fline(0.5,0,1,0,d,a )  
      \iline(1,0,2,0,d,a )  
      \fline(2,0,2.5,0,d,a )  

      \fline(0.5, 1.0, 1.5, 1.0, d, a )  
      \fline(1.5, 1.0, 2.5, 1.0, d, a )
    }{}  
    \ifthenelse{\equal{4}{#1}}
    {
      \fline(1.0, 1.0, 1.5, 0.5, d, a )  
      \fline(1.5, 0.5, 2.0, 1.0, d, a )  
      \iline(1.5, 0.0, 1.5, 0.5, d, a )  

      \iline(1.0, 1.0, 2.0, 1.0, d, a )  
      \fline(0.5, 0.0, 1.5, 0.0, d, a )  
      \fline(1.5, 0.0, 2.5, 0.0, d, a )  

      \fline(0.5, 1.0, 1.0, 1.0, d, a )  
      \fline(2.0, 1.0, 2.5, 1.0, d, a )
    }{}  
    \ifthenelse{\equal{5}{#1}}
    {
      \iline(1,0,2,1,d,a )  
      \iline(2,0,1,1,d,a )  

      \fline(0.5,0,1,0,d,a )  
      \fline(1,0,2,0,d,a )  
      \fline(2,0,2.5,0,d,a )  

      \fline(0.5,1,1,1,d,a )  
      \fline(1,1,2,1,d,a )  
      \fline(2,1,2.5,1,d,a )
    }{}  

  \end{tikzpicture}
}

\newcommand{\siteVertexA}[1]
{
  \begin{tikzpicture}[scale=.8, baseline=(current bounding box.center)]
      \iline(1,0.25,1,.75,d,a )  

      \fline(0.5,0,1,0.25,d,a )  
      \fline(1,0.25,1.5,0,d,a )  

      \fline(0.5,1,1,.75,d,a )  
      \fline(1,0.75,1.5,1,d,a )  
    \node [above, black] at (1,.8) {\tiny\ensuremath{i_1}};
    \node [below, black] at (1,.2) {\tiny\ensuremath{i_2}};
    \node [left, black] at (0.5,1) {\tiny\ensuremath{1}};
    \node [left, black] at (0.5,0) {\tiny\ensuremath{2}};
    \node [right, black] at (1.5,1){\tiny\ensuremath{1'}};
    \node [right, black] at (1.5,0){\tiny\ensuremath{2'}};
  \end{tikzpicture}
}

\newcommand{\siteVertexB}[1]
{
  \begin{tikzpicture}[scale=.8, baseline=(current bounding box.center)]
      \iline(1,0.25,1,.75,d,a )  

      \fline(0.5,0,1,0.25,d,a )  
      \fline(1,0.25,1.5,0,d,a )  

      \fline(0.5,1,1,.75,d,a )  
      \fline(1,0.75,1.5,1,d,a )  
    \node [above, black] at (1,.8) {\tiny\ensuremath{i_1}};
    \node [below, black] at (1,.2) {\tiny\ensuremath{i_2}};
    \node [left, black] at (0.5,1) {\tiny\ensuremath{1}};
    \node [left, black] at (0.5,0) {\tiny\ensuremath{2}};
    \node [right, black] at (1.5,1){\tiny\ensuremath{2'}};
    \node [right, black] at (1.5,0){\tiny\ensuremath{1'}};
  \end{tikzpicture}
}

\begin{document}
 \title{Renormalization group analysis of dipolar Heisenberg model on square lattice} 
\author{Ahmet Kele\c{s}}
\affiliation{Department of Physics and Astronomy,
  University of Pittsburgh, Pittsburgh, Pennsylvania 15260, USA}
\affiliation{Department of Physics and Astronomy,
  George Mason University, Fairfax, Virginia 22030, USA}
\author{Erhai Zhao}
\affiliation{Department of Physics and Astronomy,
  George Mason University, Fairfax, Virginia 22030, USA}

\begin{abstract}  
  We present a detailed functional renormalization group
  analysis of spin-1/2 dipolar Heisenberg model on square lattice. This model
  is similar to the well known $J_1$-$J_2$ model and describes the pseudospin
  degrees of freedom of polar molecules confined in deep optical lattice with
  long-range anisotropic dipole-dipole interactions. Previous study of this
  model based on tensor network ansatz indicates a paramagnetic
  ground state for certain dipole tilting angles which can be tuned in
  experiments to control the exchange couplings. The tensor ansatz formulated
  on a small cluster unit cell is inadequate to describe the spiral order, and
  therefore the phase diagram at high azimuthal tilting angles remains
  undetermined.  Here we obtain the full phase diagram of the model from
  numerical pseudofermion functional renormalization group calculations. We
  show that an extended quantum paramagnetic phase is realized between the N\'{e}el
  and stripe/spiral phase. In this region, the spin susceptibility flows
  smoothly down to the lowest numerical renormalization group scales with no
  sign of divergence or breakdown of the flow, in sharp contrast to the flow
  towards the long-range ordered phases.  Our results provide further evidence
  that the dipolar Heisenberg model is a fertile ground for quantum spin
  liquids.
\end{abstract} 

\maketitle
\section{Introduction}

A paradigmatic model for frustrated quantum magnetism is the $J_1$-$J_2$ model
on square lattice.  It is defined as a spin-1/2 Heisenberg model with
antiferromagnetic nearest neighbor ($J_1$) and next-nearest neighbor ($J_2$)
exchange couplings, described by the Hamiltonian
\begin{equation} H_{J_1J_2} = J_1 \sum_{\langle ij
    \rangle}\mathbf{S}_i\cdot\mathbf{S}_j + J_2 \sum_{\langle\langle ij
    \rangle\rangle}\mathbf{S}_i\cdot\mathbf{S}_j.  \end{equation}
Here the first (second) sum is over the nearest (next nearest) neighbors and
$\mathbf{S}_i=(S_i^x,S_i^y,S_i^z)$ are the usual spin-1/2 operators at site
$i$. 
Although the limits of small and large $J_2/J_1$ are well understood to have
long-range N\'eel and columnar 
orders respectively, the ground state near the
maximally frustrated regime $J_2\sim 0.5J_1$ is still controversial (see
Refs.~\onlinecite{Richter2009,Jiang2012,PhysRevB.88.060402,PhysRevLett.111.037202,
  PhysRevLett.113.027201,PhysRevB.94.075143} and references therein). There is
strong evidence that it is likely a quantum spin liquid which does not have
any conventional magnetic long range order and does not break the symmetry of
the Hamiltonian. Quantum spin liquids manifest a series of novel properties
such as topological order and excitations with fractional statistics
\cite{savary2016quantum,lee2014quantum,RevModPhys.89.025003}. They are of
great interest to strongly correlated electron systems including copper oxide
superconductors \cite{anderson1987resonating} and frustrated quantum magnets
\cite{savary2016quantum,lee2014quantum,RevModPhys.89.025003}.  
An ensuing theoretical challenge is to identify realistic physical models that
can be realized cleanly in experiments and find the parameter regions in the
phase diagram where a spin liquid arises.

\begin{figure}[h]
  \includegraphics[scale=0.8]{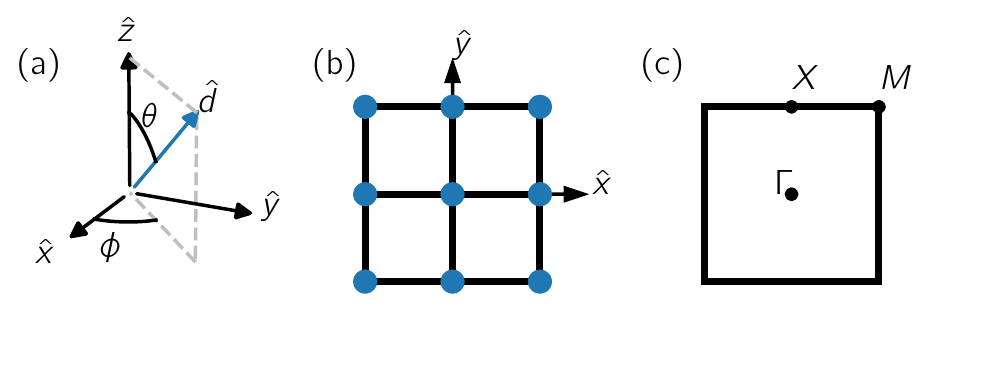}
  \caption{(Color online) Geometry of the dipolar Heisenberg model on square
    lattice.  There is one polar molecule localized on each site of the square
    lattice within the $xy$ plane (b).  Each molecule carries pseudo-spin 1/2.
    Their dipole moments are aligned along a common direction $\hat d$
    specified by the polar angle $\theta$ and azimuthal angle $\phi$ (a). The
    Brillouin zone and selected high symmetry points are shown in (c).}
  \label{fig:geometry}
\end{figure}

Recent work examined the phase diagram of dipolar Heisenberg model on square
lattice and found evidence for a possible spin liquid phase
\cite{PhysRevLett.119.050401}. The dipolar Heisenberg model can be viewed as a
close cousin of $H_{J_1J_2}$ but with a larger parameter space and important
distinctions. Its Hamiltonian is given by
 \begin{equation}
    H_{\hat d}  = 
    \sum_{i_1\neq i_2} J_{ \hat d }( i_1,i_2 ) 
    \mathbf{S}_{i_1} \cdot \mathbf{S}_{i_2}  ,
    \label{eq:hamiltonian}
\end{equation}
where the summation is over \emph{all} pairs of sites, labelled by the site
index $i_1$ and $i_2$, within the two-dimensional square lattice on the $xy$
plane (Fig. \ref{fig:geometry}). This all-to-all coupling differs from the
$J_1$-$J_2$ model. The spin exchange has the following dipolar interaction form
\begin{equation}
    J_{\hat d}( i_1,i_2 ) 
    = J_0[1-3(\hat r_{i_1i_2}\cdot \hat d)^2 ]/r_{i_1i_2}^3 
    \label{eq:dipolar_J}
\end{equation}
where $\mathbf{r}_{i_1i_2}=\mathbf{r}_{i_1}-\mathbf{r}_{i_2}$ for spins at
sites $i_1$ and $i_2$. We take the lattice constant to be unity and the energy
units such that  $J_0=1$.  The unit vector $\hat d$ is a tuning parameter of
the model (controlled by an external field), and it is conveniently
parametrized by the polar angle $\theta$ and azimuthal angle $\phi$ as shown
in Fig.~\ref{fig:geometry},
\begin{equation}
\hat d = (\sin\theta\cos\phi, \sin\theta\sin\phi, \cos\theta).
\end{equation}
Note that  the exchange $J_{\hat d}( i_1,i_2 )$ is not only long-ranged but
also anisotropic, i.e.  both its magnitude and sign of depend on the relative
orientation	 of $\hat d$ and $\hat r_{i_1i_2}$.  For example, the exchange
between two nearest neighbor spins along the $x$ direction may differ from
that along the $y$ direction, as $\hat d$ is tilted from the $z$-axis. By
tuning $\hat d$, the system may be brought to a regime that is more frustrated
than the $J_1$-$J_2$ model.

The dipolar Heisenberg model may appear foreign and artificial from a solid
state perspective.  However, it arises naturally in ultracold quantum gases of
magnetic atoms and polar molecules.  For example, as discussed in details in
Refs.
\onlinecite{yao2015quantum,PhysRevLett.119.050401,PhysRevLett.113.195302,PhysRevA.84.033619},
the pseudo-spin 1/2 describes two rotational states of the polar molecules
such as KRb confined in a deep optical lattice, the spin exchange is mediated
by the dipole-dipole interaction between the molecules, 
Eq. \eqref{eq:dipolar_J}, and $\hat d$ is the direction of all the dipoles
along an external electric field. Experiments have successfully realized the
dipolar Heisenberg model on cubic optical lattice and measured its spin
dynamics \cite{yan2013observation,bohn2017cold}.  We note that similar spin
models with long range interactions can also be realized using cold atoms with
large magnetic moments \cite{PhysRevLett.111.185305}, atoms in the highly
excited Rydberg states \cite{schauss2012observation,labuhn2016tunable}, and
trapped ions\cite{britton2012engineered,islam2013emergence}.

Previously, Zou, Liu and one of us solved $H_{\hat{d}}$ on  square lattice by
using the tensor network ansatz and keeping only the nearest and next nearest
exchange couplings \cite{PhysRevLett.119.050401}. They found evidence for a
quantum paramagnetic phase, likely a spin liquid, sandwiched between the
N\'eel and stripe phase.  The ansatz employed an $L\times L$ unit cell with
periodic boundary conditions. It was  unable to go beyond $\phi\sim 20^\circ$,
because  the small cluster cannot accommodate the incommensurate spiral order
which becomes relevant at these higher $\phi$ values.  Thus, a full phase
diagram of $H_{\hat{d}}$ from tensor networks is still lacking.  To get a
better understanding of the model, an independent method is needed. First, the
method should be able to take into account the long range interactions
faithfully and avoid severe truncations in the interaction range.  Second, it
should work directly with infinite lattice in the thermodynamic limit  to
accurately describe the spiral oder to predict the phase diagram for all
values of $\theta$ and $\phi$. Third, it should go beyond the leading order
spin wave theory \cite{PhysRevLett.119.050401} or random phase approximation
by treating all instabilities on same footing without bias. 

In this paper, we adopt a method that satisfy these three requirements above.
We obtain the zero temperature (ground state) phase diagram of $H_{\hat d}$
using the pseudo-fermion functional renormalization group analysis. We show
that the dipolar Heisenberg model shows, besides the N\'eel, stripe and spiral
phases, an extended quantum paramagnetic region where long range order is
suppressed from $\phi=0^\circ$ all the way up to $\phi=45^\circ$. This
observation is in broad agreement with previous results from different
methods. It is also in line with recent theoretical evidence of spin liquid
phase for the dipolar Heisenberg model on the triangular lattice
\cite{yao2015quantum,keles2018absence}.

The rest of the paper is organized as follows: In Sec.~\ref{sec:frg} we
present details of the pseudo-fermion functional renormalization group.  In
Sec.~\ref{sec:frg-A}, we outline the renormalization flow equations for the
single particle self-energy and two particle vertex in a compact form. In
Sec.~\ref{sec:frg-B}, we introduce the necessary parametrizations of the two
particle vertex that exploits  the symmetries of our problem to make the
numerical solution feasible.  In Sec.~\ref{sec:frg-C}, we provide details of
our numerical implementation. In Sec.~\ref{sec:results}, we present our main
results on the dipolar Heisenberg model model including the phase diagram and
the FRG flows for representative points in each phase.  The long range ordered
phases and the quantum paramagnetic phase are discussed in separate
subsections.  Finally, we summarize our main observations in
Sec.~\ref{sec:conc} and discuss their experimental implications. 

\section{Pseudo-fermion Functional renormalization group} 
\label{sec:frg}
To tackle the many-body spin problem of $H_{\hat{d}}$ in
Eq.~\eqref{eq:hamiltonian}, we first recast it in a fermionic representation
by using \begin{equation} \mathbf{S}_i=\frac{1}{2}
  \boldsymbol{\sigma}_{\alpha'\alpha} \psi\dgr_{\alpha'i}\psi_{\alpha i}
  \label{S-psi} \end{equation} similar to the parton construction used in the
study of frustrated quantum magnets and sometimes referred to as the Abrikosov
fermion representation. Here $\psi$'s are anti-commuting fermion field
operators and $\sigma$'s are the usual spin-1/2 Pauli matrices. After the
substitution, the Hamiltonian $H_{\hat d}$ becomes
\begin{equation}
    H[\bar\psi,\psi]  = \sum_{i_1i_2}\frac{J_{i_1i_2}}{4} 
    \boldsymbol{\sigma}_{\al_1'\al_1}\cdot \boldsymbol{\sigma}_{\al_2'\al_2}
    \psi\dgr_{\al_1' i_1}
    \psi\dgr_{\al_2' i_2}\psi_{\al_2 i_2}
    \psi_{\al_1 i_1}.
    \label{eq:hamiltonian_psi}
\end{equation}
We will drop the subscript $\hat d$ for $H$ in the rest of the paper for
brevity.  This Hamiltonian for fermions has quartic spin-dependent
interactions but no hopping between sites. Thus the bare single-particle Green
function is only frequency dependent (the chemical potential is kept at zero
throughout the calculation)
\begin{equation}
  G^{(0)} (\omega) = \frac{1}{i\omega}
\end{equation}
which comes from imaginary time derivative term in the action
$S[\bar\psi,\psi] =\bar\psi \partial_\tau\psi+H[\bar\psi,\psi]$. The
translation of the spin problem to a fermion problem enables one to use the
well-established many-body techniques for correlated electrons to understand
the ground state of the system. Note, however, that the fermion problem Eq.
\eqref{eq:hamiltonian_psi} is very peculiar: the interaction energy is much
larger (in fact infinitely larger) than the kinetic energy. For this reason,
we resort to functional renormalization group which is capable of describing
such strongly interacting models
\cite{Metzner2012,P.KopietzL.BartoschSchutz2010}. 

Functional renormalization group (FRG) is an elegant theoretical framework
that implements the Wilsonian scale transformation in a systematic way to
integrate out the high energy degrees of freedom and obtain a low energy
effective field theory.  There are several alternative functional
renormalization techniques suitable for the Hamiltonian in
Eq.~\eqref{eq:hamiltonian_psi}. Here, our main goal is to provide an impartial
diagnosis of competing phases and the many-body instabilities at low energies.
To this end, we employ a purely fermionic FRG scheme without auxiliary
Hubbard-Stratanovich fields. This approach is known as pseudo-fermion
functional renormalization group (pf-FRG) \cite{Reuther2010} and it is proven
successful in identifying spin liquid behavior in a variety of models
\cite{Reuther2010,Reuther2011,Reuther2011b,Reuther2014,Buessen2016,Iqbal2016}.

\subsection{Flow Equations} 
\label{sec:frg-A}
The starting point of pf-FRG is the fermionic renormalization flow equation
derived from vertex expansion up to one loop order:
\begin{align}
    \frac{d}{d\Lambda}
    \Sigma( \omega_1 ) 
    &=-\sum_2 \Gamma_{1,2;1,2} S(\omega_2)  ,
    \label{eq:self_energy_flow}
    \\
    \frac{d}{d\Lambda}
    \Gamma_{ 1',2';1,2 } 
    &= \sum_{3,4} \Pi(\omega_3,\omega_4 ) 
    \biggr[
     \frac{1}{2} \Gamma_{ 1',2';3 ,4 }\Gamma_{ 3 ,4 ;1 ,2 } 
     \nonumber\\
    &-\Gamma_{ 1',4 ;1 ,3 }\Gamma_{ 3 ,2';4 ,2 } 
    +\Gamma_{ 2',4 ;1 ,3 }\Gamma_{ 3 ,1';4 ,2 }
    \biggr],
    \label{eq:vertex_flow}
\end{align}
where we leave scale ($\Lambda$) dependence implicit in the self energy
$\Sigma$, the two particle vertex $\Gamma$, the full propagator $G$, and the
single-scale propagator $S$ for brevity.  The subscripts are shorthand
notation,  for example, 
\begin{equation}
\Gamma_{1',2';1,2}\equiv\Gamma(
i_1',\alpha_1',\omega_1',i_2',\alpha_2',\omega_2';
i_1 ,\alpha_1 ,\omega_1 ,i_2 ,\alpha_2 ,\omega_2  )
\nonumber
\end{equation}
with site index $i$, spin $\alpha$, and frequency $\omega$ (we only consider
zero temperature so the Matsubara frequency becomes continuous variable). The
summation denotes integration over continuous frequencies and summation over
lattice sites and spin. The two scale-dependent propagators defined by
$G(i',\w',\al';i,\w,\al)=\delta_{i'i}\delta_{\w,\w'}\delta_{\al',\al}G(\w)$
and
$S(i',\w',\al';i,\w,\al)=\delta_{i'i}\delta_{\w,\w'}\delta_{\al',\al}S(\w)$
are diagonal in site, spin and frequency space at all stages of
renormalization where
\begin{equation}
G(\w)=
\frac{\Theta(|\omega|-\Lambda)}{i\omega+\Sigma(\omega)},\quad
S(\w)=
\frac{\delta(|\omega|-\Lambda)}{i\omega+\Sigma(\omega)},
\label{eq:propagators}
\end{equation}
and $\Sigma(\w)$ is the self-energy.  By using a diagrammatic expression
for the vertex 
\begin{equation}
  \Gamma_{1',2',1,2} =\vertex{1,2,1',2'} ,
  \nonumber\\
\end{equation}
Eq.~\eqref{eq:vertex_flow} can be represented diagrammatically by the familiar
particle-particle, particle-hole and exchange channels as shown by the
following one-loop diagrams:
\begin{equation}
    \oneloop{1,2,1',2',3,4}+
\oneloop{1,1'/,2/,2',3,4/}+
\oneloop{1,2'/,2/,1',3,4/}
\nonumber
\end{equation}
Different from the usual practice of FRG applied to correlated
electrons~\cite{P.KopietzL.BartoschSchutz2010}, pf-FRG uses a modified
expression for the product of Green functions (polarization bubbles) by using
the following \emph{full} derivative
\begin{equation}
 \Pi(\omega_3,\omega_4 ) =-\frac{d}{d\Lambda}
 \left[
 G(\omega_3)G(\omega_4)
 \right].
\end{equation}
which includes terms beyond one-loop expansion \cite{PhysRevB.70.115109}. 

The expressions given in Eq.~\eqref{eq:self_energy_flow} and
\eqref{eq:vertex_flow} forms a non-linear integro-differential system of
equations, with the initial condition defined by the bare Hamiltonian $H_{\hat
  d}$ at the ultraviolet (UV) scale ($\Lambda_\mathrm{UV}\rightarrow\infty$)
  such that
\begin{align}
  \Sigma(\omega)\big|_{\Lambda=\Lambda_\mathrm{UV}}   &= 0,
  \label{eq:initial_condition}\\
  \Gamma_{1,2;1',2'}\big|_{\Lambda=\Lambda_\mathrm{UV}} &= \mathcal{\hat A} \frac{1}{4}
  \sigma^\mu_{\alpha_1\alpha_1'} \sigma^\mu_{\alpha_2\alpha_2'}
  J_{i_1,i_2}\delta_{i_1i_1'}\delta_{i_2i_2'} 
    \nonumber
\end{align}
where $\mathcal{\hat A}$ is the antisymmetrization operator.  The flow
equations for the two-particle vertex and the single particle self-energy
describe ordering tendencies as the RG scale $\Lambda$ is systematically
lowered from $\Lambda_\mathrm{UV}$.  

\subsection{Parametrization of the Vertex} 
\label{sec:frg-B}
To reduce the computational cost, we use the symmetries of the system such as
spin $SU(2)$ invariance, translational invariance, and the lattice point group
symmetry. These symmetries can be taken into account in an efficient way by
using suitable parametrizations of the two particle vertex.  For example, we
perform the lattice parametrization by using
\begin{equation}
    \Gamma_{1',2';1,2} = \mathcal{\hat A} 
      \Gamma_{12}^{1',2',1,2}
    \delta_{i_1'i_1} \delta_{i_2'i_2}.
    \label{eq:site_param}
\end{equation}
Here we have shortened the notation such that the lower indices stand for
sites $i$ and the upper indices stand for the frequency and spin $(\w,\alpha)$
as follows
\begin{align}
\mathcal{O}_{12}&\rightarrow\mathcal{O}(i_1,i_2)\nonumber\\
\mathcal{O}^{1',2',1,2}&\rightarrow\mathcal{O}(\al_1',\w_1',\al_2',\w_2',\al_1,\w_1,
\al_2,\w_2).
\end{align}
Note that full vertex $\Gamma_{1',2;,1,2}$ is distinguished from the
site-parametrized vertex $\Gamma_{12}^{1',2',1,2}$ by its arguments. 
The parametrization Eq. \eqref{eq:site_param} can be expressed diagrammatically
\begin{equation}
  \vertex{1,2,1,2}  = 
  \delta_{i_1'i_1} \delta_{i_2'i_2} \siteVertexA{1,2,3,4,5,6} 
 - \delta_{i_1'i_2} \delta_{i_2'i_1} \siteVertexB{1,2,3,4,5,6} 
 \nonumber
\end{equation}
where the site parametrized vertex is depicted by a zigzag line. After
substitution of Eq. \eqref{eq:site_param} in the flow equation
\eqref{eq:vertex_flow} and equating the terms associated with
$\delta_{i_1'i_1}\delta_{i_2'i_2}$ and $\delta_{i_1'i_2}\delta_{i_2'i_1}$
separately we find
\begin{widetext} 
\begin{equation}
  \frac{d}{d\Lambda}\Gamma_{12}^{1'2'12} =
  \sum_{\w_3\al_3,\w_4\al_4}\Pi(\w_3,\w_4) \biggr[ 
  \underbrace{ \Gamma_{12}^{1'2'34} \Gamma_{12}^{3412} }_{\oneloopV{1}}
   - \underbrace{ \sum_{i_3}\Gamma_{13}^{1'413}\Gamma_{32}^{32'42}}_{\oneloopV{2} }   
  +\underbrace{\Gamma_{12}^{1'413} \Gamma_{22}^{32'24}}_{\oneloopV{3}}  
  +\underbrace{\Gamma_{11}^{1'431}\Gamma_{12}^{32'42}}_{\oneloopV{4}} 
  +\underbrace{\Gamma_{12}^{42'13}\Gamma_{12}^{1'342}}_{\oneloopV{5}} 
  \biggr]
\end{equation}
\end{widetext}
where the diagrammatic expression of each term is presented below an
underbrace. These five diagrams are well known in many-body theory: the first
term is the particle-particle ladder whereas the last diagram is the
particle-hole ladder.  The third and fourth diagrams are vertex corrections.
The second diagram is the RPA bubble. A key strength of the pf-FRG approach is
that the parametrization of the two particle vertex Eq. \eqref{eq:site_param}
along with the full propagators Eq. \eqref{eq:propagators} enforces exactly
one fermion per site throughout the flow. This ensures that Eq. \eqref{S-psi}
is a faithful representation of the original spin problem (empty or double
occupation of any site is strictly forbidden).  The preservation of fermion
number constraint within pf-FRG has been numerically demonstrated in
Ref.~\onlinecite{Buessen2017} (see also Ref.~\onlinecite{Reuther2010} for more
details).

Finally, we use the following spin parametrization for systems with $SU(2)$
symmetry
\begin{align}
    \Gamma_{12}^{1'2'12} &= 
    \Gamma_{i_1i_2}^\mathrm{spin}(\w_1',\w_2',\w_1,\w_2)
    \boldsymbol{\sigma}_{\al_1'\al_1}\cdot\boldsymbol{\sigma}_{\al_2'\al_2}  \nonumber\\
    &+\Gamma_{i_1i_2}^\mathrm{dens}(\w_1',\w_2',\w_1,\w_2)
    \delta_{\al_1'\al_1}\delta_{\al_2'\al_2}.  
\end{align}
This leads to two distinct sets of coupled flow equations for
$\Gamma^\mathrm{spin}$ in the spin channel and $\Gamma^\mathrm{dens}$ in the
density channel. The resulting equations are rather lengthy and can be found
in Ref.~\onlinecite{Reuther2010}. The vertex functions are expressed in terms
of Mandelstam variables $s=\w_1+\w_2=\w_1'+\w_2'$, $t=\w_1-\w_3$,
$u=\w_1-\w_4$ by a change of variable. These variables efficiently encode the
symmetries in frequency space such as frequency conservation
$\w_1+\w_2=\w_1'+\w_2'$.  We also exploit the reflection symmetry with respect
to the plane containing the dipole direction $\hat{d}$ and perpendicular to
the square lattice. Note that the $C_4$ rotational symmetry is broken once the
dipoles are tilted, $\theta\neq 0$.

\subsection{Numerical Implementation} 
\label{sec:frg-C}

Translational invariance implies that the vertex functions $\Gamma_{i_1i_2}$
only depend on the distance between sites $i_1$ and $i_2$. In our numerics, we
use the translational invariance to fix $i_1$  as a reference site and
consider all $\Gamma_{i_1i_2}^\mathrm{spin}$ and
$\Gamma_{i_1i_2}^\mathrm{dens}$ with $i_2$ within an $N_L\times N_L$ square
region centered at $i_1$. Formally this scheme corresponds to an infinite
system with finite truncation of interaction range. (In the pf-FRG literature,
this is sometimes referred as ``the cluster size'' for brevity.) The frequency
$\w$ is discretized and lives on a logarithmic frequency grid of $N_\omega$
points from a very large ultraviolet scale $\Lambda_\mathrm{UV}$, down to a
very small infrared scale $\Lambda_\mathrm{IR}$. Since the only energy scale
of the problem is the dipolar exchange $J_0=1$, the small and large energy
cutoffs should satisfy $\Lambda_\mathrm{UV}\gg1$ and
$\Lambda_\mathrm{IR}\ll1$, respectively.  We slowly reduce the RG scale
$\Lambda$ all the way from the ultraviolet to the infrared with four steps
between two consecutive frequencies on the grid.  This gives $N_\Lambda=4N_\w$
renormalization steps in total. We solve the first order coupled flow
equations \eqref{eq:self_energy_flow}-\eqref{eq:vertex_flow} with the initial
condition \eqref{eq:initial_condition} for the dipolar Heisenberg model
\eqref{eq:hamiltonian} on the multidimensional grid of lattice sites and
frequencies described above using the fourth order Runge-Kutta algorithm.  The
overall computational cost of the numerical solution scales with
$N_\Lambda\cdot N_L^2\cdot N_\omega^4$. We perform simulations for lattice
sizes up to $N_L=11$ and $N_\omega=64$ frequencies and for $N_\Lambda$ that
takes up to 8 renormalization steps between neighboring frequencies. We
checked that increasing these parameters does not change our results
significantly for several selected points in the phase diagram.  Our code is
written in Python to run on Graphical Processing Units (GPU) with massive
parallelism implemented by using the open source Numba compiler. As an
example, for the system sizes mentioned above, a single simulation for a given
set of parameters takes about 4.5 hours in a state-of-the-art GPU such as
NVIDIA TITAN Xp with 3840 cuda cores.

Typically, there are about 20-30 million running couplings ($\Gamma$'s) being
monitored during each step of the FRG flow. Analysis of such a large
collection of coupling constants is facilitated by the calculation of certain
two particle correlation functions. At each RG scale, the single-particle
self-energy and the two-particle interaction vertex can be used to obtain the
static spin susceptibility in real space using
\begin{eqnarray} 
  \chi_{i_1,i_2} &=& \int_0^\infty d\tau \langle T
  S_{i_1}(\tau) S_{i_2}(0) \rangle \\ 
  &=& 
  \begin{tikzpicture}[scale=.5,
    baseline={([yshift=-.1cm]current bounding box.center)}]
      
      \draw[densely dashed,very thick] (0,0.7) -- (1,0.7); 
      
      \draw[middlearrow={stealth}] 
      (1,0.7) to [out=75,in=105] (3,.7) ;
      \draw[middlearrow={stealth reversed}] 
      (1,0.7) to [out=-75,in=-105] (3,.7) ;
    
      \draw[densely dashed, very thick] (3,0.7) -- (4,0.7); 
      
      \node[above,black] at (4.5,0.2) {$+$};
      
      \draw[densely dashed, very thick] (5,0.7) -- (6,0.7); 
      
      \draw[middlearrow={stealth}] 
      (6,0.7) to [out=85,in=125] (8.0,0.9) ;
      \draw[middlearrow={stealth reversed}] 
      (6,0.7) to [out=-85,in=-125] (8.0,0.5) ;
    
      \draw[fill=gray] (8,0.5) rectangle (8.4,.9);
     
      \draw[middlearrow={stealth}] 
      (8.4,0.9) to [out=55,in=95] (10.4,0.7) ;
      \draw[middlearrow={stealth reversed}] 
      (8.4,0.5) to [out=-55,in=-95] (10.4,0.7) ;
     
      \draw[densely dashed, very thick] (10.4,0.7) -- (11.4,0.7);
    \end{tikzpicture}, \nonumber 
\end{eqnarray}
where ``$\begin{tikzpicture} \draw[densely dashed, very thick] (0,0) --
  (.5,0); \end{tikzpicture} $'' is the diagrammatic representation for the
Pauli matrix $\sigma$, related to the spin at site $i$ by
$S_i=1/2\sigma_{\alpha'\alpha} \psi\dgr_{\alpha'i}\psi_{\alpha i} $. After
Fourier transforming to the momentum space, the spin susceptibility
$\chi(\mathbf{p})$ gives clues to the leading ordering instabilities, if any,
as the renormalization flow approaches the infrared scale.  For example, the
locations of the susceptibility maxima $\chi_\mathrm{max}$ in the Brillouin
zone determine the ordering wave vector for the incipient long range order.
Typically $\chi_\mathrm{max}$ displays a Curie-Weiss-like behavior at large RG
scales $\Lambda \gg 1$. The effects of quantum correlations start to emerge
around $\Lambda \approx 1$. If there is an instability toward long ranged
order, below a critical scale $\Lambda_c<1$, the susceptibility shows rapid
increase until the flow breaks down and is replaced with unphysical jumps.  On
the other hand, the susceptibility may continuously flow to lowest numerical
renormalization scale $\Lambda\rightarrow \Lambda_\mathrm{IR}$. This points to
a quantum paramagnetic phase such as a spin liquid.
Examples of these different flow behaviors will be given below.

\section{Phase diagram from FRG} 
\label{sec:results}

\begin{figure}
  \includegraphics[width=0.45\textwidth]{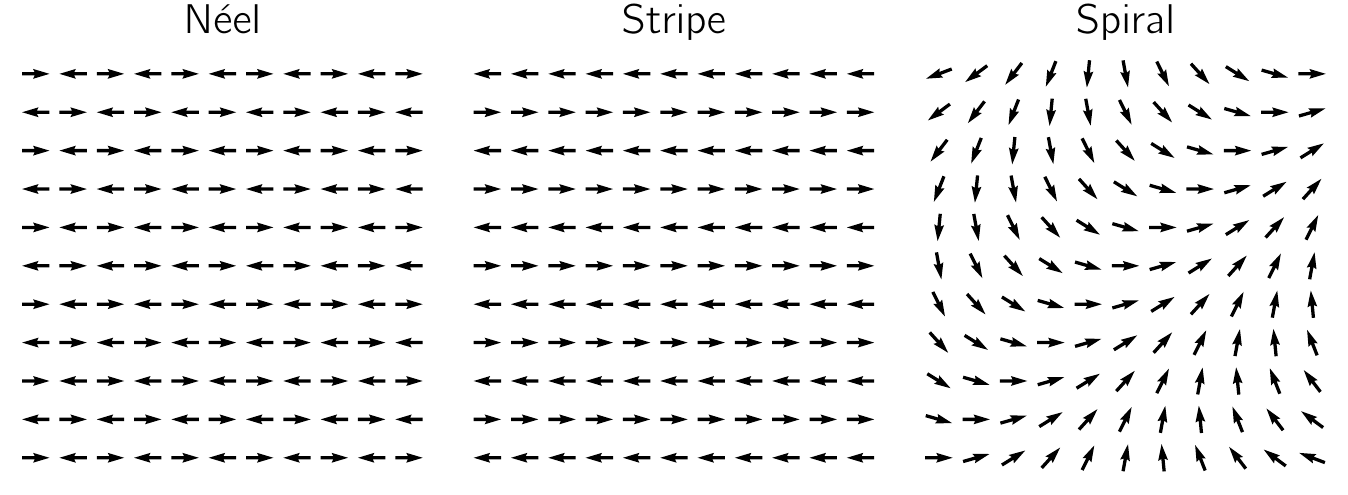}
  \caption{Spin configurations for the N\'eel, stripe and spiral order. These
    are the three competing classical long-range orders for the dipolar
    Heisenberg model on square lattice.} 
  \label{fig:spin_config}
\end{figure}

Before discussing the full FRG results, we first review the classical limit of
the dipolar Heisenberg model on square lattice, previously discussed in Ref.
\onlinecite{PhysRevLett.119.050401}.  The classical phase diagram contains
three phases schematically shown in Fig. \ref{fig:spin_config}.  The N\'eel
order is stabilized for small values of $\theta$. It gives way to the stripe
order at large $\theta$ if $\phi$ is not too large.  For large $\phi$ close to
$45^\circ$ and large $\theta$, the system is in the spiral phase.  FRG
provides an elegant way to obtain the classical phase diagram via the solution
of the flow equations by ignoring all frequency dependences.  This method has
been shown to be consistent with random phase approximation and
Luttinger-Tisza method \cite{PhysRevB.96.045144}. It also serves as a useful
benchmark for FRG.  Specifically, we start from  the UV scale with the initial
condition Eq. \eqref{eq:initial_condition} and numerically monitor the flows
of the frequency-independent vertices under the sliding renormalization scale
$\Lambda$. When the absolute maximum of the vertex reaches a large cutoff
value, a divergence is detected and we stop the flow.  The scale at which this
cutoff value is reached gives us the critical ordering scale $\Lambda_c$,
which can be interpreted roughly as an estimation of the critical temperature.
The corresponding classical order is found by Fourier transforming the
susceptibility and examining the location of its peaks. 

The resulting critical scales are shown in false color in the top row of
Fig.~\ref{fig:square}.  Here the color yellow (blue) indicates high (low)
values of the critical scale $\Lambda_c$.
The contour lines give a rough guide for the phase boundaries (not shown
explicitly to avoid clutter).  The three classical phases show up as three
plateaus of $\Lambda_c$ in the parameter space of the $\theta-\phi$ plane.
For the antiferromagnetic N\'eel order at small dipolar tilting $\theta$, the
susceptibility shows four maxima at the corners of the Brillouin zone (the
$M$-point, see Fig.~\ref{fig:geometry}). As $\theta$ is increased, peaks at
the corners of the Brillouin zone start to extend and eventually merge at the
$X$-point.  For larger $\theta$, the susceptibility peak moves to the
$X$-point, indicating the stripe order. With $\theta$ fixed but increasing the
azimuthal angle $\phi$ beyond a critical value, the peaks at the $X$ points
start moving towards the $\Gamma$ point, the center of the Brillouin zone,
indicating the spiral order.  From Fig.~\ref{fig:square}, we see that the
stripe order typically has large critical scales whereas $\Lambda_c$ is
suppressed close to the N\'eel-stripe phase boundary. The suppression is most
severe near a region around $\theta\sim \phi\sim 45^\circ$.  In the next
subsection, we analyze the FRG flow equations with full frequency dependence.
Special attention will be given to regions where the long range orders are
suppressed.

\subsection{Three long-range ordered phases} 

\begin{figure}
  \includegraphics[scale=1]{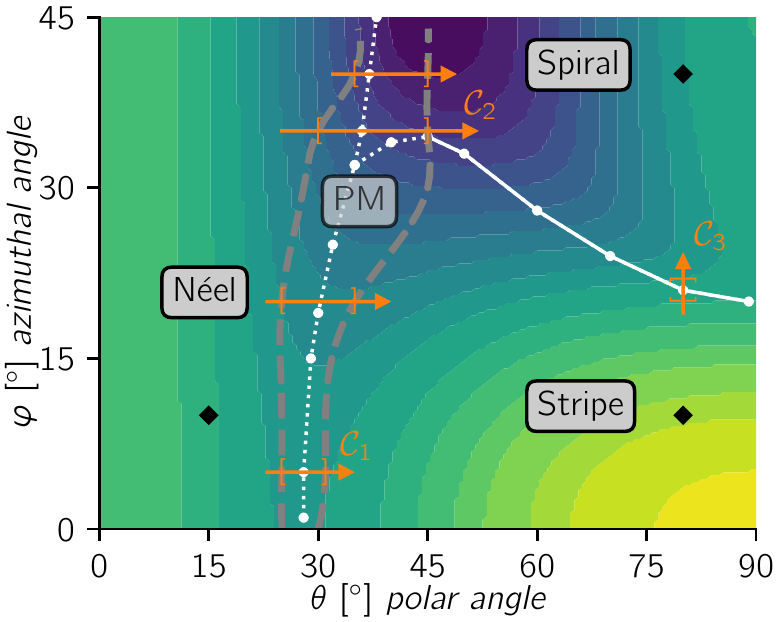}
  \includegraphics[scale=1]{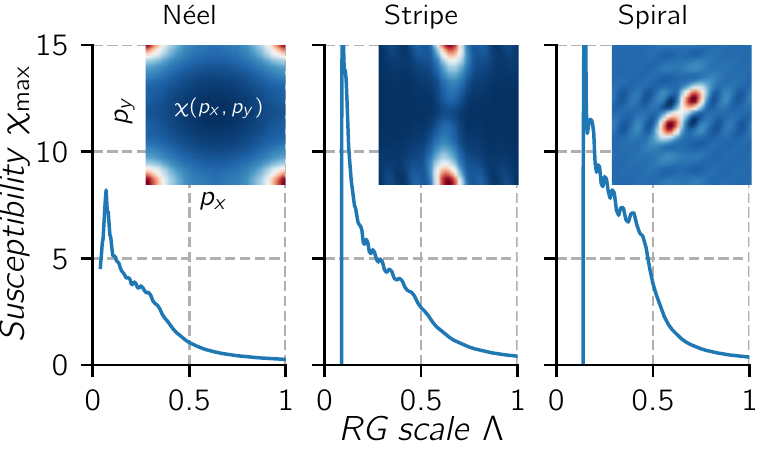} 
  \caption{(Color online) Phase diagram of dipolar Heisenberg model on square
    lattice as function of dipole tilting angles $\theta$ and $\phi$. Three
    long-range orders are N\'eel, stripe and spiral phases.  Color contours
    are critical RG scales of obtained from frequency independent FRG. For
    three selected points (black diamonds) in the phase diagram, the flows of
    maximum susceptibility are shown in the lower panel. The corresponding
    susceptibility profiles throughout the Brillouin zone at a small RG scale
    are shown as the insets of the lower panel. The orange arrows denoted by
    $\mathcal{C}_1$, $\mathcal{C}_2$ and $\mathcal{C}_3$ are cuts near the
    phase boundaries. The dashed lines indicate the estimated boundary of a
    quantum paramagnetic (PM) phase.
  }
  \label{fig:square} 
\end{figure}

The main results of our full FRG calculations are summarized in Fig.
\ref{fig:square}.  We systematically perform multiple cuts through the
$\theta-\phi$ plane and examined the spin susceptibility profiles in the
momentum space in conjunction with the renormalization flow of
$\chi_\mathrm{max}$ to determine the many-body ground state.  The resulting
phase boundaries are shown in Fig.~\ref{fig:square} with white lines, overlaid
on top of the false color $\Lambda_c$ obtained from frequency independent FRG
discussed above. The lower panel shows the RG flows of $\chi_\mathrm{max}$ for
a selected point from each phase. The insets show the corresponding
susceptibility profiles within the Brillouin zone. The region between the grey
dashed lines in the vicinity of the phase boundaries is another phase and it
will be described separately in the next subsection.

\begin{figure*}
  \centering 
  \includegraphics[scale=1]{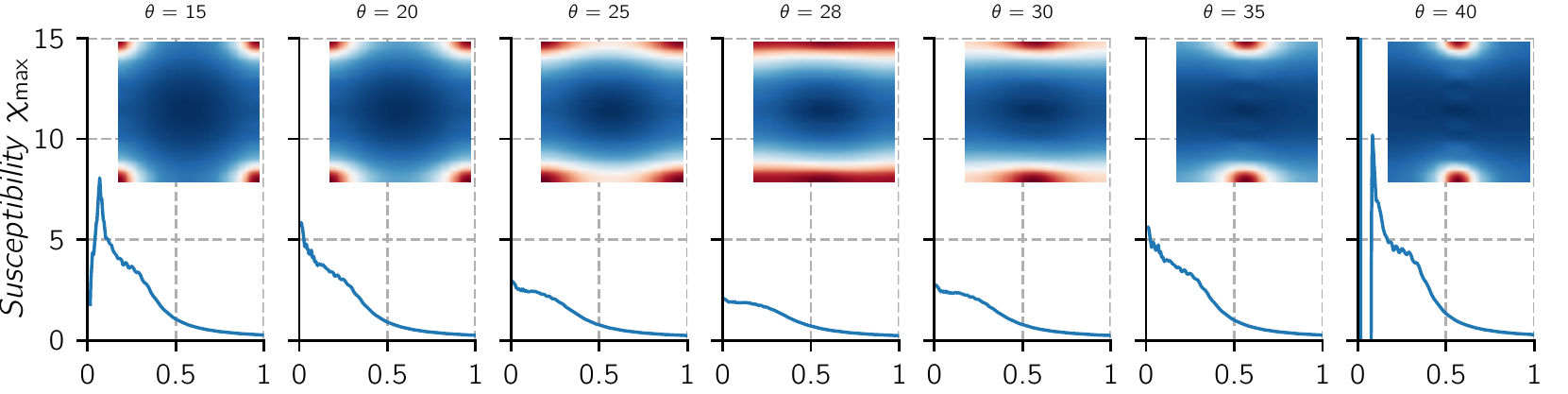}
  \includegraphics[scale=1]{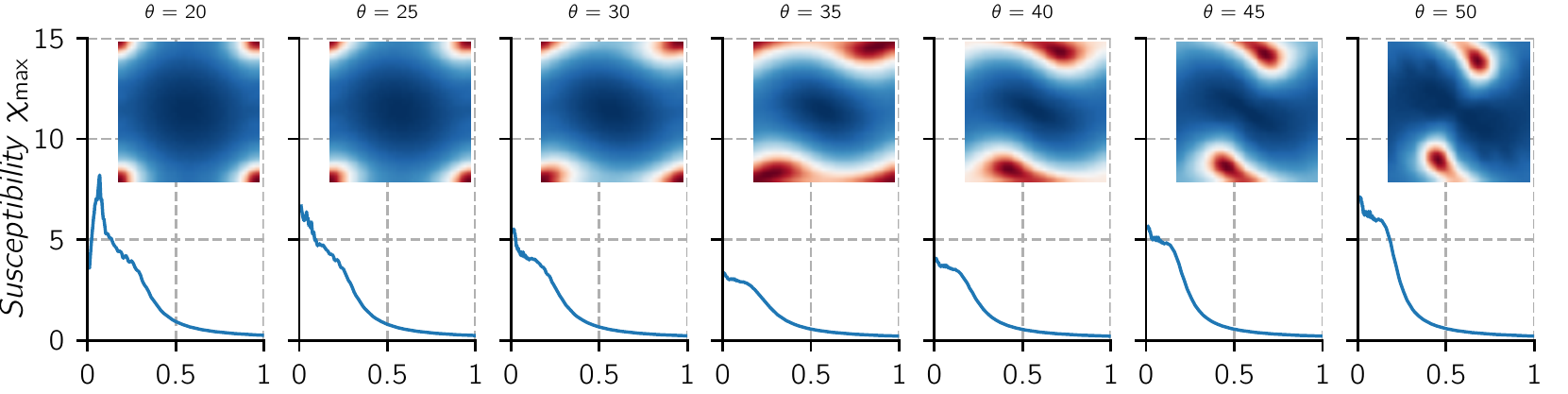}
  \includegraphics[scale=1]{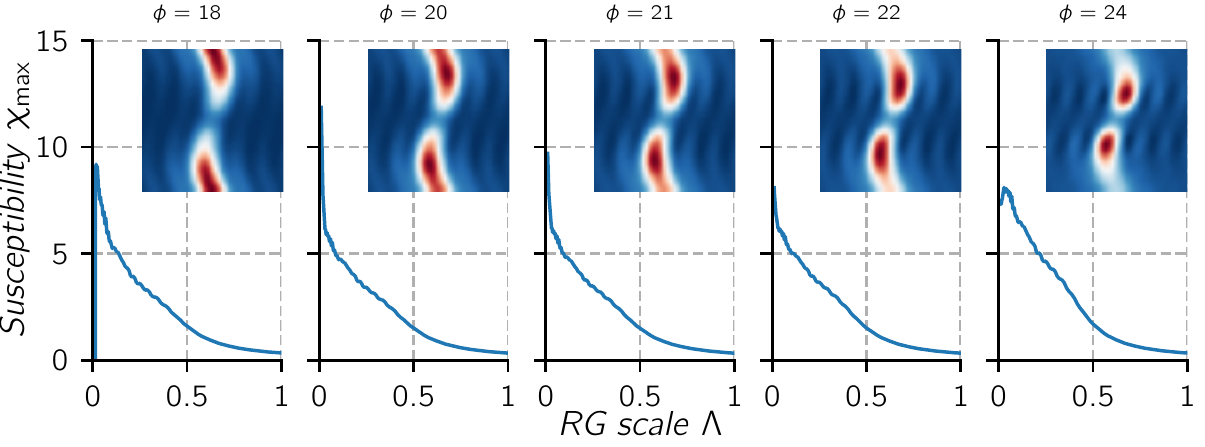} 
  \caption{(Color online) Renormalization group flows of maximum
    susceptibility $\chi_\mathrm{max}$ along three cuts in the vicinity of the
    phase boundaries. Top row is the $\mathcal{C}_1$ cut from the N\'eel phase
    to stripe phase, middle row is the $\mathcal{C}_2$ cut from the N\'eel
    phase to spiral phase and the bottom row is the $\mathcal{C}_3$ cut from
    the stripe to spiral phase. Insets show the susceptibility profiles near
    the end of each flow.  Within an extended window of angle $\theta$ (top
    and middle row), the susceptibility flows smoothly down to
    $\Lambda\rightarrow 0$ and shows no sign of divergence or breakdown. This
  points to a paramagnetic ground state.} 
  \label{fig:cuts}
\end{figure*}

The full FRG predicts three long range ordered phases. We can understand each
phase by selecting a representative point (the black diamonds $\bdiamond$ in
Fig.~\ref{fig:square}) in the phase diagram  and inspecting its numerical FRG
data.  Let us start with $\theta=15^\circ$ and $\phi=10^\circ$,  a point deep
inside the N\'eel phase. The spin susceptibility profile $\chi(\pb)$ over the
full Brillouin zone at a small RG scale $\Lambda\approx 0.4$ (shown in the
inset of lower panel) clearly indicates the leading spin correlations are of
N\'eel type because of the peaks at the corners of the Brillouin zone.
However, the peak position by itself is not sufficient to identify the
presence of a complete instability. The susceptibility data as a function of
the RG scale should also be inspected. To this end, we focus on
$\chi_\mathrm{max}$ at the peak position, the $M$-point. Its renormalization
flow is shown in the lower panel of Fig.~\ref{fig:square}. Here, as $\Lambda$
is gradually reduced, we first observe an upturn followed a shoulder with tiny
oscillations around $\Lambda\sim0.2-0.3$. These oscillations are due to the
discretization in the frequency grid. They are well controlled, and can be
reduced by using a finer grid. 
Upon further decreasing $\Lambda$, a steep increase of the susceptibility is
observed indicating a divergence being developed. 
Shortly afterward, however, the continuous flow breaks down and is replaced by
unphysical, discontinuous evolution of $\chi$ (not shown for lower values of
$\Lambda$).
The breakdown of smooth pf-FRG flow is in a large part due to the finite
truncation $N_L$ of the effective interaction range in our numerical
implementation. The truncation regulates the divergence and eventually leads
to unphysical flows at low $\Lambda$.  A faithful description of the
divergence would require diverging correlation length, i.e. ever increasing
$N_L$.
Even though a true divergence is hard to reach in finite $N_L$ implementation
of pf-FRG, one can make sure the flow indeed suggests long range order by
systematically varying $N_L$. 
In practice, the breakdown of the continuous flow is a clear indication of
incipient long range order in pf-FRG provided that $N_L$ is sufficiently large
(see Ref.~\onlinecite{Reuther2010} and Ref.~\onlinecite{keles2018absence} for
a detailed discussion).

Similar results are shown in Fig.~\ref{fig:square} for two points deep inside
the stripe and spiral phases respectively. For the stripe phase at
$\theta=80^\circ$, $\phi=10^\circ$, the susceptibility peak is at the
$X$-point signaling the alternating layered structure shown in the middle
panel of Fig.~\ref{fig:spin_config}. The spiral phase at $\theta=80^\circ$,
$\phi=40^\circ$ has an ordering wave vector corresponding to the
incommensurate spin texture as shown in the right panel of
Fig.~\ref{fig:spin_config}. Note that the susceptibilities flow up to much
larger values in the stripe and spiral phases in compared to the N\'eel phase,
since the selected points in these phases are further away from the phase
boundary.

We can determine the boundary between these long range ordered phases by
tracking the peak positions of the susceptibility.  Because the peak becomes
broadened near the phase boundaries, it is much easier to monitor the
degeneracy of $\chi_\mathrm{max}$ and define the phase boundary as where it is
most degenerate, i.e., the peak is most smeared and extended. This yields the
dotted and solid white lines in Fig. \ref{fig:square}.  One can check these
lines are exactly where the peak position changes qualitatively, for example,
from peak at $M$ (N\'eel order) to peak at $X$ (stripe order). The solid white
line gives an accurate phase boundary between the stripe and the spiral phase.
On the other hand, we emphasize that the dotted white line separating the
N\'eel and stripe/spiral phase is {\it not} the physical phase boundary. In
the next subsection, we show that an extended quantum paramagnetic phase is
sandwiched between these phases.
 
\subsection{A robust quantum paramagnetic region} 

Now we show that within a rather broad region near the N\'eel-stripe and the
N\'eel-spiral phase boundary, enclosed by the dashed lines in
Fig.~\ref{fig:square}, long range order is suppressed, and the spin
susceptibility flows smoothly and continuously down to the lowest RG scale
$\Lambda_\mathrm{IR}$ without any indication of a divergence being developed.
Thus the ground state within this region is a quantum paramagnet according to
the FRG.  The quantum paramagnetic region spans a width of about $5^\circ$ to
$15^\circ$ in the $\theta$ direction, and persists to all $\phi$ values.

To demonstrate the paramagnetic behavior in the vicinity of the phase
boundary, we take several cuts indicated by the orange arrows in
Fig.~\ref{fig:square}.  The detailed FRG flows are shown in
Fig.~\ref{fig:cuts} for three typical cuts labeled by $\mathcal{C}_1$ to
$\mathcal{C}_3$. Among these, $\mathcal{C}_1$ is a cut from the N\'eel phase
going into the stripe phase at $\phi=5^\circ$, $\mathcal{C}_2$ is a cut from
the N\'eel phase to the spiral phase at $\phi=35^\circ$, and $\mathcal{C}_3$
is a cut from the stripe to the spiral phase with fixed $\theta=80^\circ$ but
increasing $\phi$. Along the $\mathcal{C}_1$ cut (top row of
Fig.~\ref{fig:cuts}), the flow pattern in the beginning (e.g.
$\theta=15^\circ$) clearly indicates a long range N\'eel order. As we increase
the dipolar tilting $\theta$, the sharp peaks at $M$ become broadened and
extend towards each other along the $MX$ line. At the same time, the
development of divergence in the maximum susceptibility $\chi_\mathrm{max}$ at
low $\Lambda$ is gradually weakened.  At $\theta=28^\circ$, the peak at $M$ is
no longer visible. The spin susceptibility reaches maximum along the entire
line connecting $M$ and $X$.  Here, the RG flow of $\chi_\mathrm{max}$ remains
remarkably smooth down to lowest numerical RG scale without any sign of
instability, see the top row, middle panel of Fig.~\ref{fig:cuts}. With
further increase in $\theta$, the susceptibility develops a new peak at $X$,
and the flow of $\chi_\mathrm{max}$ becomes divergent at low $\Lambda$ again,
signaling the stripe order. The width of the quantum paramagnetic region is
estimated to be about $5^\circ$ for the $\mathcal{C}_1$ cut. The rough
boundary of the paramagnetic region is indicated by the orange square bracket
markers in Fig.~\ref{fig:square}. 

The $\mathcal{C}_2$ cut (the middle row of Fig.~\ref{fig:cuts}) also reveals a
similar  quantum paramagnetic region. But this time, the region is
significantly larger, within a window about $15^\circ$ wide (see the orange
bracket in  Fig.~\ref{fig:square}).  Another cut above $\mathcal{C}_2$
indicates that the paramagnetic region narrows down at larger $\phi$ values.
So the maximum quantum paramagnetic region is located where all three phases
meet, around the $\mathcal{C}_2$ cut. 

We also checked whether any quantum paramagnetic behavior persists near the
stripe to spiral phase boundary. The $\mathcal{C}_3$ cut (bottom row of
Fig.~\ref{fig:cuts}) reveals that the susceptibility flow in this region is
markedly different compared to the $\mathcal{C}_1$ and $\mathcal{C}_2$ cuts.
The flows indeed become smooth down to $\Lambda_\mathrm{IR}$ in a very narrow
window of about $2^\circ$ in $\phi$, but they are not qualitatively different
from nearby points along the cut. In particular, there is no clear suppression
of susceptibility as observed in $\mathcal{C}_1$ and $\mathcal{C}_2$ cuts.
Therefore, along along the $\mathcal{C}_3$ cut, a direct transition from
stripe order to spiral order is observed, with no clear evidence for an
intermediate paramagnetic phase with appreciable width. 

\subsection{Comparison with other methods}

The same model $H_{\hat{d}}$ has been investigated previously using tensor
network ansatz in Ref. \onlinecite{PhysRevLett.119.050401}, where a quantum
paramagnetic phase with a width of about 1 to 2 degrees was found for $\phi$
from $0^\circ$ up to $20^\circ$. Beyond $\phi\sim 20^\circ$, the tensor
network algorithm becomes inaccurate due to the small cluster size which is
incompatible with the spiral order. For this reason the phase boundary for
$\phi>20^\circ$ is not known from the tensor algorithm.  The dipolar
Heisenberg model was also solved by spin wave analysis and Schwinger boson
mean field (SBMF) theory in Ref. \onlinecite{PhysRevLett.119.050401}.  Both
methods predicted a spin liquid region between the N\'eel and stripe phase for
$\phi$ up to $\sim 35^\circ$, but the exact shape and position of the spin
liquid phase are different. For example, SBMF yields a wider liquid region
(the method is known to have the tendency of overestimating disordered
phases). Finally we emphasize that in Ref.
\onlinecite{PhysRevLett.119.050401}, the exchange couplings $J_{ij}$ are
truncated, i.e., only the nearest and next-nearest neighbor exchanges are
retained.

The FRG approach adopted here is very different from these previous methods.
For example, it does not work directly with variational wave functions or
order parameter fields, and focuses instead on the correlation functions under
the RG flow.  Despite the difference, FRG also predicts a quantum paramagnetic
region separating the N\'eel order and the stripe order, in broad agreement
with Ref. \onlinecite{PhysRevLett.119.050401}. 
Taken together,  these numerical evidences consistently point to a quantum
paramagnetic phase in the dipolar Heisenberg model on square lattice.
The width of the paramagnetic region predicted from FRG is larger than that
from tensor networks. We believe this is mainly due to that fact that longer
range exchanges are kept in FRG, i.e. $J_{ij}$ for $|i-j|\leq N_L\gg 1$, which
lead to stronger frustration and a more robust paramagnetic ground state
compared to the $J_1$-$J_2$ model.
It is also interesting to compare FRG with the modified spin wave theory which
contains the leading terms in the $1/S$ expansion, as well as SBMF which can
be related to the large $N$ limit of $Sp(N)$ spin models.  The perturbative
diagrammatic expansions in the three methods are rather different.
The detailed analysis of FRG for spin-$S$ and $SU(N)$ spin models can be found
in Refs. \onlinecite{PhysRevB.96.045144, buessen2017functional}.

The new insight from our FRG calculation is that the paramagnetic region will
persist to higher $\phi$ values, all the way to $\phi=45^\circ$. FRG works
with infinite lattice and a large cutoff $N_L$ of the effective interaction
range, and therefore is much better equipped to describe the spiral order.
Near the classical N\'eel-spiral phase boundary, both orders are very weak
with the critical temperature $T_c$ significantly suppressed (see for example
the dark region in the top row of Fig. \ref{fig:square}). They are melted by
quantum fluctuations to form a quantum paramagnetic ground state.  It is
challenging to precisely determine the phase boundary between the paramagnetic
phase and the long range order phases in pf-FRG.  The dashed lines in Fig.
\ref{fig:square} are results of a conservative estimation, and the quantum
paramagnetic phase may actually occupy a larger region in the phase diagram.
We hope our results can stimulate further work with large scale numerics and
different methods to shed more light on this intriguing region. 

\section{Conclusion} 
\label{sec:conc}

Our main result is that a quantum paramagnetic phase occupies an extended
region in the phase diagram of  $H_{\hat d}$ on square lattice thanks to the
long-range anisotropic dipolar exchange. Recall that in the $J_1$-$J_2$ model,
finite $J_2$ leads to exchange frustration, and magnetic order is suppressed
for $J_2/J_1\sim 0.5$.
Here longer range exchange couplings tend to amplify the frustration.  And
tilting the dipoles alone is sufficient to achieve the frustration needed for
a quantum paramagnetic phase.
Dipole tilting also break the four-fold rotational symmetry of $H_{\hat d}$ to
favor the spiral order at large $\theta$ and $\phi$.  The paramagnetic phase
is most robust (i.e. has the largest expanse in parameter space) in regions
where all three long orders meet and compete, near the $\mathcal{C}_3$ cut in
Fig. \ref{fig:square}. 
The melting of magnetic orders as a consequence of dipolar exchange coupling
is a general phenomenon. It has also been demonstrated for $H_{\hat d}$ on the
triangular lattice  \cite{yao2015quantum,keles2018absence} and kagome lattice
\cite{yao2015quantum}.

In conclusion, we have demonstrated via numerical functional renormalization
group that the spin-1/2 dipolar Heisenberg model is an excellent candidate for
studying frustrated magnetism and searching for quantum spin liquids. Such
spin models with long range dipolar exchange has already been realized in
experiments with ultracold KRb molecules in deep optical lattices, and the
spin dynamics has been measured by microwave spectroscopy
\cite{yan2013observation}.  We hope further progress in cooling the molecular
gases \cite{wu2012ultracold} and obtaining lattice fillings close to unity
\cite{moses2015creation} can enable direct observation and measurement of the
phase diagram of the dipolar Heisenberg model.

\begin{acknowledgments}
We thank Johannes Reuther for illuminating discussions on FRG, and W. Vincent
Liu and Haiyuan Zou for sharing their insights about the tensor network ansatz
and spin wave analysis. This work is supported by NSF PHY-1707484 and AFOSR
Grant No.  FA9550-16-1-0006. A.K. also acknowledges support from ARO Grant No.
W911NF-11-1-0230.
\end{acknowledgments}

\bibliography{refs}

\end{document}